\documentclass[10pt,letterpaper,twocolumn]{article} 

\usepackage{ol}
\usepackage{amsmath}

\begin{document}

\twocolumn[ 

\title{Spectral-discrete solitons and  localization in frequency space}

\author{A.V. Gorbach and D.V. Skryabin}
\address{Centre for Photonics and Photonic Materials,
Department of Physics, University of Bath, Bath BA2 7AY, UK}

\begin{abstract}
We report  families of discrete optical
solitons in frequency space, or spectral-discrete solitons
existing in a dispersive Raman  medium,
where individual side-bands are coupled by 
coherence.  The associated time-domain
patterns correspond to either trains of ultrashort pulses, or weakly modulated waves.
We describe the physics behind the spectral localization and
study soliton bifurcations, stability and dynamics.
\end{abstract}

] 

Discrete  solitons in systems of coupled waveguides
have been a subject of intense recent interest, due to the richness of
optical effects and  the potential applications associated with them \cite{chris,kivshar}.
A seemingly different, and also very active subfield of optics is the generation
of ultrashort pulses via the excitation of multiple harmonics in Raman-active
and other materials \cite{sokolov}. In this work, we reveal and explore a surprising
link between discrete optical solitons and short pulse generation in a Raman medium.

It can be seen from numerical results \cite{sokolov1,yavuz}
that the spectral harmonics involved in short-pulse
generation by Raman coherence, can either be spread over the entire frequency space
covered by the model equations, or be accidentally localized around some central frequency.
The effect of spectral localization has not attracted much
attention so far and it is  often assumed that cascaded generation of Raman
side-bands is practically limited by only the frequency dependence of Raman gain or material  losses.
Below, we neglect the above two effects and  demonstrate
the existence of spectrally localized structures supported
by a balance between the normal group velocity dispersion (GVD) of a Raman medium
(which, quite surprisingly, plays the role of a holding potential)
and the nonlinear coupling between the Raman side-bands (playing
the role of the effective discrete  diffraction happening
in the frequency space). These localized structures bear some features
of discrete solitons in waveguide arrays \cite{chris,kivshar},
but at the same time have many unique properties. It appears natural
to call them spectral-discrete solitons (SDSs).
Similar ideas of exploring localization in Fourier space have been recently
used in a different physical context
for an  interpretation of the Fermi-Pasta-Ulam paradox \cite{flach}.

We start with the well-established model \cite{sokolov,sokolov1,yavuz}
describing the evolution of $2N+1$  side bands in a Raman medium coherently
excited away from resonance. In the
dimensionless form, the corresponding equations are
\begin{eqnarray}
\label{eqE}
&& i\partial_{z} E_n=\beta_n E_n + q^* E_{n-1}+ q E_{n+1},
\end{eqnarray}
Here $n=-N,\dots,0,\dots, N$, $z$ is the dimensionless propagation coordinate and
$E_n$ are the amplitudes of the side-bands
with frequencies $\omega_n=\omega_0+n(\omega_R+j\mu)$.
$\omega_R$ is the Raman transition frequency and $j\mu$ is the
detuning of the frequency difference of the neighboring harmonics from $\omega_R$.
It has been convenient for us to split this detuning into its sign $j=\pm 1$
and its absolute value $\mu\ge 0$. $\beta_n$ is the propagation constant at the frequency $\omega_n$.
The frequency dependence of the propagation constant $\beta(\omega)$
can be fitted with a polynomial in $\omega$, which in the discretized frequency space is
\begin{equation}
\label{disp}
\beta_n=\sum_{k=0}^{K}p_k n^k.
\end{equation}
where $p_1$ is associated with the group velocity at $n=0$,
$p_2$ with the group velocity dispersion, and all the higher order
coefficients correspond to the higher order dispersions.

$q$ is the Raman coherence responsible for coupling
between the side-bands and ensuring their cascaded
excitation. Any possible frequency dependence of the coupling is
neglected for simplicity. $q$ is expressed as a product of the
amplitudes $a_{1,2}$ ($|a_1|^2+|a_2|^2=1$) of the states separated by $\omega_R$:
$q=2a_1a_2^*$. The normalized  Hamiltonian
for the two-photon Raman process (with the decay and Stark terms omitted)
is given by \cite{sokolov,sokolov1,yavuz}
\begin{equation}
\hat H=-{1\over 2}\left(\begin{array}{cc} j\mu & S\\S^* & -j\mu \end{array}\right),
~S=\sum_n E_nE_{n+1}^*.\label{hamilt}
\end{equation}
If $S$ varies slowly in comparison to the separation of the $\hat H$
eigenvalues, then from
$i\partial_t\vec a=\hat H\vec a$, where $\vec a=(a_1,a_2)^T$, 
an approximate expression for $q$ can be found:
\begin{eqnarray}
\label{stacQ}
q&=&{jS \over 2\sqrt{\mu^2+|S|^2}},
\end{eqnarray}
where $j=+1/-1$ corresponds  to the in-phase/anti-phase dressed
state, respectively \cite{sokolov,yavuz}.

If $q$ is real, the correspondence between Eqs.~(\ref{eqE}) and the equations
describing propagation in an array of coupled linear waveguides \cite{chris,kivshar} is obvious:
$q$ plays the role of the coupling between the waveguides.
However, in our case the coupling  itself is a nonlinear
function of the field amplitudes, see Eqs. (\ref{stacQ}), (\ref{hamilt}). The combination of
equations (\ref{eqE}) and (\ref{stacQ}) is invariant with respect to the symmetry transformation
\begin{equation}
E_n\to E_n e^{in\psi},\label{phase}
\end{equation}
where $\psi$ is an  arbitrary constant.
Thus the phase difference between the neighboring side-bands
is an arbitrary number, which does not change  the
properties of any solution found. For $q$ real, its sign, $s=sign(q)$, is controlled
by the relative phases of the Raman harmonics. The in-phase neighboring harmonics correspond to
$\psi=0$ and $s=1$, while the out-of-phase harmonics imply that $\psi=\pi$ and $s=-1$. However,
the overall sign of $q^*E_{n-1}+qE_{n+1}$ is not changed under the transformation (\ref{phase})
with any $\psi$, including $\psi=\pi$. This is
because the phase of $q$, see Eq. (\ref{stacQ}), compensates the phases of $E_{n\pm 1}$.
It contrasts sharply with the   discrete waveguide array model  \cite{kivshar}, where
Eq. (\ref{phase}) is not a symmetry, and therefore,
taking $\psi=\pi$ reverses the discrete diffraction sign, which
strongly effects propagation in the array \cite{chris,kivshar}.

It is also useful to consider the continuous limit with $E_n \to E(\omega)$,
$\beta_n \to \beta(\omega)$, and
$E_{n\pm1} \to E(\omega)\pm \delta\omega\cdot \partial_{\omega} E+{1\over 2}
(\delta\omega)^2 \cdot\partial_{\omega}^2 E + O(\delta\omega^3)$,
where $\delta\omega=(\omega_R+j\mu)/\omega_0$.
In this limit, Eqs.~(\ref{eqE}), (\ref{stacQ})
transform into the Schr\"odinger equation for $E(\omega)$:
\begin{eqnarray}
&& i\partial_{z} E = \beta(\omega)E + j|q|\frac{\partial^2 E}{\partial \omega^2}+2j|q|E,
\label{eqE_cont}\\
&& |q|={{\cal P}\over 2\sqrt{\mu^2+{\cal P}^2}},~
 {\cal P}=\int \left|E(z,\omega)\right|^2 d\omega.\label{qI}
\end{eqnarray}
the $\beta(\omega)$-term forms a potential in frequency space,
while $q$ is the coefficient in front of the effective diffraction term controlling
the spreading of a wavepacket in frequency space.  $\cal P$ is the total power,
which is a conserved quantity ($\partial_z{\cal P}=0$)
and so in the continuous limit, $q$ itself is a constant. The third term
in (\ref{eqE_cont}) is simply a constant shift of the propagation constant.
We note that  $j=-1$ corresponds to the sign in front of the second derivative being that which is usual
in the quantum mechanical context. Therefore, if $\beta(\omega)$ has a minimum at
some $\omega$ it indicates the presence of a potential well in frequency space.

The simplest parabolic  potential well is formed by $p_2>0$ and $p_{k>2}=0$,
corresponding  to the normal GVD typical for
Raman  gases \cite{sokolov}.
It is clear that by taking account of GVD only, i.e. $p_{k>2}=0$,
we reduce Eq. (\ref{eqE_cont})
to the  harmonic oscillator problem, which has a known set of localized eigenmodes.
Thus  Eqs. (\ref{eqE}), (\ref{stacQ}) are also expected to have a set of localized solutions.
In a way, this is analogous to the approach used for fiber solitons. Their
existence was first discovered theoretically in the approximation of
zero higher-order dispersions, and later their relative robustness to
the realistic dispersion profiles  was verified \cite{kivshar}.
To find 
stationary
solutions to the nonlinear system (\ref{eqE}), (\ref{stacQ})
we make the ansatz
\begin{equation}
\label{sol_E}
E_n(z)=B_n\exp(i\kappa z+i\alpha n z)
\end{equation}
and numerically solve the resulting algebraic equations for $B_n$ using a Newton method.
$\kappa$ and $\alpha$ are the two  parameters enabled by
phase symmetries of our system.  We notice that substitutions
$\kappa\to \kappa-p_0$ and $\alpha\to \alpha-p_1$
eliminate parameters $p_{0,1}$. In other words, without restriction of generality
one can assume $p_{0,1}=0$ in Eq. (\ref{disp}).

\begin{figure}
\begin{center}
\includegraphics[angle=270,width=0.45\textwidth]{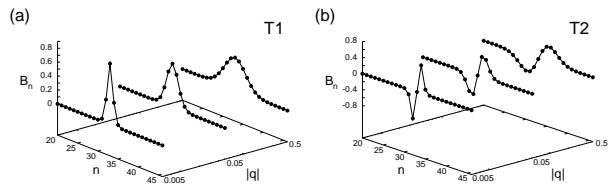}
\end{center}
\caption{Single-peak (ground state) and two-peak SDSs calculated for
$j=-1$. $\kappa$ varies from $-0.004$ to $-0.93$ in (a) 
and from $0.009$ to $-0.79$ in (b), which gives $|q|$
from $0.005$ to $0.5$. $p_2=0.01$, $\mu=1$, $\alpha=0$, $p_{k>2}=0$.}
\end{figure}

\begin{figure}
\begin{center}
\includegraphics[angle=270,width=0.35\textwidth]{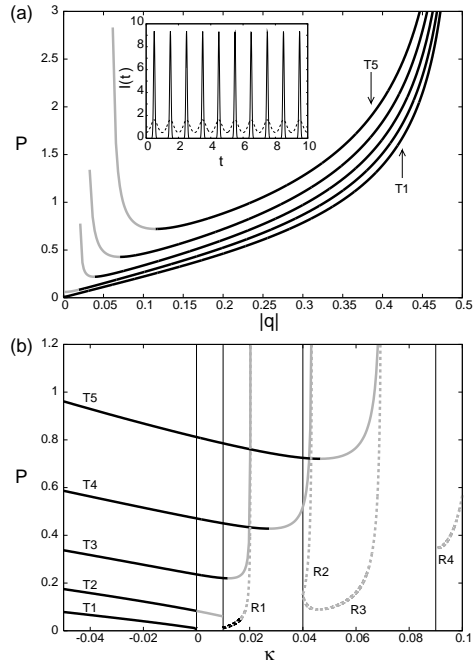}
\end{center}
\caption{Solid lines show power of the first five SDSs vs $|q|$ (a) and
vs $\kappa$ (b) for $j=-1$. Dashed lines in (b) show $j=+1$ SDSs.
The other parameters as in Fig. 1. Black/gray
lines mark stable/unstable solutions.}
\end{figure}

Using the above approach we have been able to find families of spectrally
localized solutions, which we called spectral-discrete solitons (SDSs).
Examples of SDSs corresponding to the first and second
modes of the harmonic oscillator are shown in Figs. 1(a), (b) as functions
of the coherence $|q|$. 
Doing numerical continuation of SDSs along $\kappa$
we  computed $q$ at each step and used $|q|$ as a parameter in Figs. 1,2.
$|q|$ tends to its maximum, $|q|=1/2$, when $|S|\gg 1$.
Notably, the derivative of $|q|$ in $|S|$ tends to zero under the same conditions,
i.e. $|q|$ becomes field independent.
 Thus, perhaps counter-intuitively,  in the limit of the maximum Raman coherence
Eqs. (\ref{eqE}) become quasi-linear. Oppositely, the nonlinearity of our system
is at its maximum for $|S|\ll 1$.
The dependencies of the power $P=\sum_n|E_n|^2$
of the first five SDSs (labeled as $T1$-$T5$ in all Figs.) 
as functions of both $|q|$ and $\kappa$ are shown in Figs. 2(a,b).
Black lines in Fig. 2 correspond to the stable SDSs and gray ones to the unstable ones.
The boundaries of existence in  $\kappa$ of some of the SDSs
(see vertical lines in Fig. 2(b)) are given by the linear spectrum of Eqs. (\ref{eqE}).

$|q|$ tends to $1/2$, when $\kappa$ is negative and its absolute value increases.
For these parameter values, the shape of SDSs is very close to the
modes of the linear harmonic oscillator. Since  nonlinearity is
diminished under these conditions, it is not surprising that SDSs are stable for $\kappa<0$.
Diffraction in the frequency space increases for large $|q|$, and so is the number
of excited Raman harmonics. Thus in the time domain, the quasi-linear SDSs
represent a periodic sequence of ultra-short pulses, see the full line in the inset to
Fig.~2(a), where the intensities of the total field $I=\left|
\sum_n E_n(z)\exp(i\omega_n t)\right|^2$ are plotted. Here, $t$ is dimensionless time.
At intermediate and low values of Raman coherence, the diffraction in frequency
space is relatively small and nonlinear
interaction starts to play an important role.
Both of the above factors contribute to stronger spectral
localization of SDSs, see Fig. 1. The corresponding temporal
profiles transform into weakly modulated
wavetrains, see the dashed curve in the inset to Fig. 2(a).
Dynamical instabilities of the higher order SDSs are readily found in this regime, see Figs. 2(a,b).
However, the ground state solution remains  stable within the entire range
of its existence. An example of evolution of spectral components of the unstable two-peak SDS
perturbed by noise is shown in Fig. 3(a). One can see that localization
due to potential $\beta(\omega)$
remains unaffected by the instability development. Shape deformation
of the unstable SDS  increasing  its power leads to
formation of spectral breathers, see Fig. 3(b).
Note that the dependencies of $P$ on $\kappa$ in Fig. 2(a)
indicate that transition from the stable to unstable regimes may be
driven by the Vakhitov-Kolokolov type of instability \cite{kivshar}.

\begin{figure}
\begin{center}
\includegraphics[angle=270,width=0.45\textwidth]{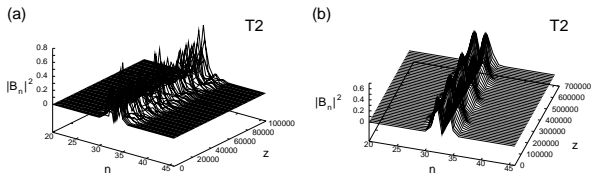}
\end{center}
\caption{Propagation of the unstable two-peak SDS with noise added (a)
and  with its power increased by a regular correction (b). $\kappa=0.005$ and $|q|\approx 0.013$.
The other parameters as in Fig. 1.}
\end{figure}

\begin{figure}
\begin{center}
\includegraphics[angle=270,width=0.22\textwidth]{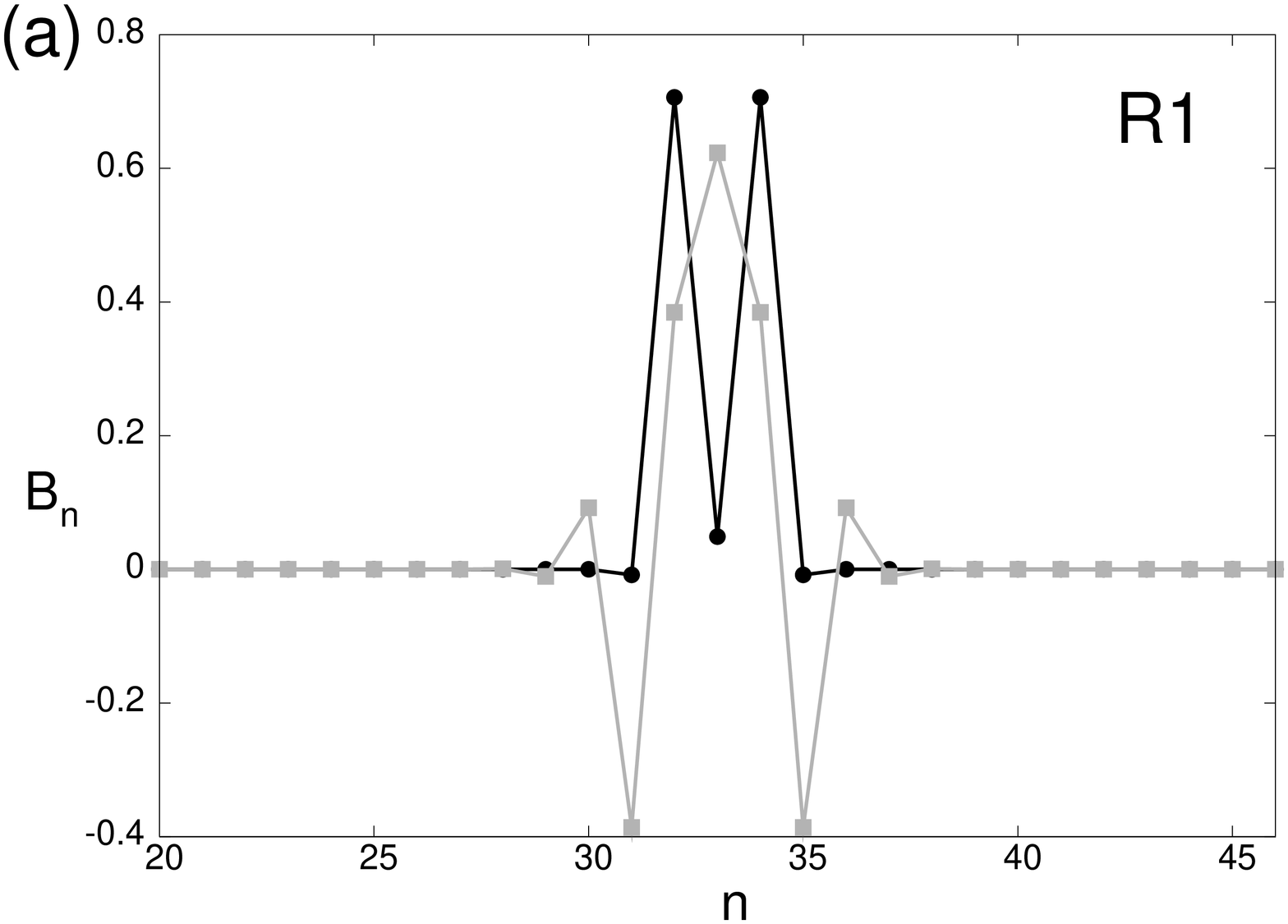}
\includegraphics[angle=270,width=0.22\textwidth]{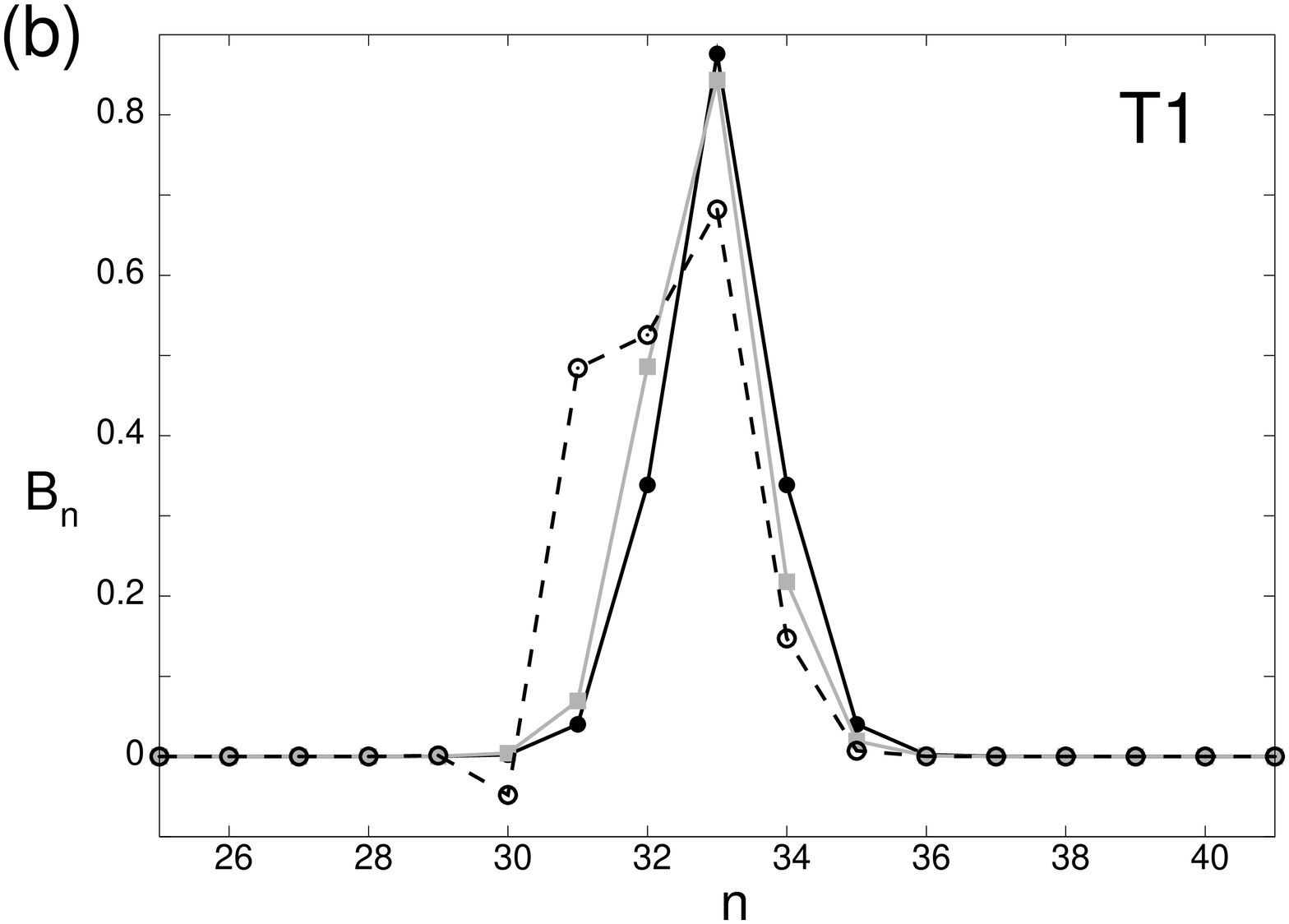}
\end{center}
\caption{(a) An example of 
SDS for $j=+1$.
$\kappa=0.01$, $|q|\approx 0.00035$ (black lines and circles),
$\kappa=0.02$, $|q|\approx 0.016$ (gray lines and squares).
(b) Examples of transformation of a single peak $j=-1$ SDS
for non-zero $p_3=0.005$ (dashed line, open circles) and
non-zero $\alpha=0.008$ (solid gray line, gray squares).
Solid black line and filled circles mark the $p_3=\alpha=0$
case. $\kappa=-0.004$ for all solutions. The other parameters as for Fig. 1.
}
\end{figure}

The above discussion focuses on the SDSs branches
found for negative Raman detuning $j=-1$ ($T$-type SDS), while the dashed lines in Fig. 2(b)
mark the solutions existing for $j=+1$ ($R$-type SDS). The latter ones are mostly unstable,
because for $j=+1$ the discrete diffraction is not compensated by the
normal GVD. However, $R$-type SDSs are still of some interest  because they are involved with the
bifurcations joining them to the $T$-type SDSs. An example of an $R$-type
solution is shown in Fig. 4(a). We have also found that the SDS branches shown in Fig. 2
can be smoothly continued using the Newton method into the region of
nonzero higher order dispersions and nonzero values of  $\alpha$, which results in the appearance of
asymmetries in their profiles, see Fig. 4(b). Interestingly, this creates an opportunity
to compensate asymmetries due to unavoidable $p_3$ with controllable $\alpha$.
A more detailed discussion of the existence and stability of SDSs goes beyond our
present scope.

Summary: We reported families of spectral-discrete solitons
existing due to the balance between the effective diffraction in frequency space
induced by Raman coherence,  and the spectral trapping
created by the normal GVD. The SDS concept has allowed
us to study the problem of short pulse generation
using  methods of nonlinear dynamics.
We demonstrated that spectrally broad, but still localized,
solutions associated with trains of short-pulses are typically stable
and that solutions strongly localized in frequency space, corresponding to the
weakly modulated waves, tend to be dynamically unstable.
Our approaches could be useful in the analysis
of other nonlinear systems which generate regular frequency combs.



\end{document}